\documentclass[nhess]{copernicus}

\begin{document}

\title{Identification of earthquake precursors in the hydrogeochemical and geoacoustic data for the Kamchatka peninsula by flicker-noise spectroscopy}

\author[1]{G. V. Ryabinin}
\author[2]{Yu. S. Polyakov}
\author[3]{V. A. Gavrilov}
\author[4,5]{S. F. Timashev}

\affil[1]{Geophysical Survey, Kamchatka Branch of the Russian Academy of Sciences, Petropavlovsk-Kamchatsky, Russia}
\affil[2]{USPolyResearch, Ashland, PA, U.S.A.}
\affil[3]{Institute of Volcanology and Seismology, Far Eastern Branch of the Russian Academy of Sciences, Petropavlovsk-Kamchatsky, Russia}
\affil[4]{Institute of Laser and Information Technologies, Russian Academy of Sciences, Troitsk, Moscow Region, Russia}
\affil[5]{Karpov Institute of Physical Chemistry, Moscow, Russia}

%% The [] brackets identify the author to the corresponding affiliation, 1, 2, 3, etc. should be inserted.

\runningtitle{Earthquake precursors in the data for the Kamchatka peninsula}

\runningauthor{G. V. Ryabinin et al.}

\correspondence{Yuriy S. Polyakov\\ (ypolyakov@uspolyresearch.com)}

\received{}
\pubdiscuss{} %% only important for two-stage journals
\revised{}
\accepted{}
\published{}

%% These dates will be inserted by the Publication Production Office during the typesetting process.

\firstpage{1}

\maketitle

\begin{abstract}
A phenomenological systems approach for identifying potential precursors in multiple signals of different types for the same local seismically active region is proposed based on the assumption that a large earthquake may be preceded by a system reconfiguration (preparation) at different time and space scales. A nonstationarity factor introduced within the framework of flicker-noise spectroscopy, a statistical physics approach to the analysis of time series, is used as the dimensionless criterion for detecting qualitative (precursory) changes within relatively short time intervals in arbitrary signals. Nonstationarity factors for chlorine-ion concentration variations in the underground water of two boreholes on the Kamchatka peninsula and geacoustic  emissions in a deep borehole within the same seismic zone are studied together in the time frame around a large earthquake on October 8, 2001. It is shown that nonstationarity factor spikes (potential precursors) take place in the interval from 70 to 50 days before the earthquake for the hydrogeochemical data and at 29 and 6 days in advance for the geoacoustic data.
\end{abstract}

%% only used for copernicus2.cls
%%\abstract{
%%TEXT
%% \keywords{TEXT}}

\introduction
%% \introduction[modified heading if necessary]
Earthquake prediction in the time frame of several months to less than an hour before the catastrophic event, which is often referred in literature as "short-term" prediction, has been a subject of extensive research studies and controversial debates both in academia and mass media in the past two decades \citep{Gel97a,Gel97b,Wys97a,Uye09a,Cic09a}. One of the key areas in this field is the study of earthquake precursors, physical phenomena that reportedly precede at least some earthquakes. The precursory signals are usually grouped into electromagnetic, hydrological/hydrochemical, gasgeochemical, geodetic, and seismic  \citep{Gel97a,Har05a,Uye09a,Cic09a}. Electromagnetic precursory signals are further classified into the signals believed to be emitted from within focal zones, such as telluric and magnetic field anomalies, and radio waves over epicentral regions \citep{Uye09a}. The localized changes in electric and magnetic fields that reportedly accompany some seismic events span a wide range of frequencies, including ULF, VLF, ELF and RF fields, and were observed in the time frame from 2-3 years to dozens of minutes prior to an earthquake \citep{Cic09a,Uye09a}. Hydrological/hydrochemical precursory signals include water level or quality changes in the weeks, days, or hours prior to a number of earthquakes, groundwater temperature changes, and variations in the concentrations of dissolved ions like chlorine or magnesium usually in the time frame of months to days before an earthquake \citep{Har05a,Cic09a,Du10a}. Gasgeochemical precursory signals comprise numerous anomalous gas emission observations, the majority of which were reported for the concentration of radon gas in the earth \citep{Har05a,Cic09a}. More than 100 studies show that changes in radon exhalation from the earth's crust precede a number of earthquakes by months, weeks, or days \citep{Cic09a}. Geodetic signals mostly include surface deformations (tilts, strains, strain rate changes) over distances of tens of kilometers that precede some major earthquakes by months to days \citep{Cic09a}. Seismic precursory signals encompass foreshocks that typically take place less than 30 days before the main shock and high-frequency (acoustic emission) and very low-frequency precursory signals that are not detected by conventional seismographs \citep{Ihm94a,Rea99a,Gor08a,Gav08a}. Another promising type of possible precursory signals is anomalous animal behavior for very short time frames (within 2-3 days, usually hours) prior to a large seismic event \citep{Kir00a,Yok03a,Li09a}.

Despite the large number of earthquake precursors reported in literature, most of which are summarized by \citet{Har05a,Cic09a}, an International Commission on Earthquake Forecasting for Civil Protection concluded on October 2, 2009, "the search for precursors that are diagnostic of an impending earthquake has not yet produced a successful short-term prediction scheme" \citep{ICE09a}. The reports of the International Association of Seismology and Physics of the Earth's Interior contain similar findings \citep{Wys97b}. The lack of confidence can be attributed to several reasons. First, some fundamental aspects of many non-seismic signals, for example, lithosphere-atmosphere-ionosphere coupling and propagation of high-frequency electromagnetic signals in the conductive earth, are unresolved, and many of the proposed physical models are questionable \citep{Uye09a}. Second, the experimental data on precursory signals are often limited to few earthquakes and few measurement sites, they frequently contain gaps and different types of noise \citep{Har05a,Cic09a,Uye09a}. Third, different techniques of identifying the anomalies are used for different signals or even in different studies for the same signal. In some cases, the anomalous changes are determined by analyzing the signals themselves \citep{Har05a,Uye09a,Cic09a}, while in other cases they are identified by studying the derived statistics or functions, such as Fisher information or scaling parameters \citep{Tel09a,Tel09b}. Moreover, seasonal changes and instrumentation or other background noise often need to be filtered out prior to the identification of precursors. 

In view of the above three problems, we believe that earthquake precursor research can be advanced by employing a phenomenological systems approach to the analysis of signals of different types in the same local geographic region. We assume that a large earthquake may be preceded by a system reconfiguration (preparation) at different time and space scales, which manifests itself in qualitative changes of various signals within relatively short time intervals. For example, such anomalous hydrogeochemical signals may be observed months to weeks before the impending earthquake, anomalous geoacoustic emissions - only days prior to the event, and anomalous behavior of animals - only hours before the catastrophe. In order to test this approach and identify different signals that may be related to a specific large seismic event, one needs to have a standard criterion or a set of standard criteria to detect signal anomalies in virtually arbitrary signals. In this study, we will use a nonstationarity factor introduced within the framework of flicker-noise spectroscopy (FNS), a statistical physics approach to the analysis of time series \citep{Tim07a,Tim07b,Tim10a}. This dimensionless criterion is practically independent from the individual features of source signals and is designed to detect abrupt structural changes in the system generating the signal, which makes it a promising candidate to be one of the standard criteria. The nonstationarity factor was previously used to detect precursors in electrochemical and telluric signals recorded in the Garm area, Tajikistan prior to the large 1984 Dzhirgatal earthquake \citep{Des03a,Vst05a}, geoelectrical signals at station Giuliano, Italy prior to several 2002 earthquakes \citep{Tel04a}, and ULF geomagnetic data at Guam prior to the large 1993 Guam earthquake \citep{Hay06a,Ida07a}. Other approaches to identifying precursory features in earthquake- and volcano-related signals, which are based on different monlinear analysis techniques, were discussed by \citet{Tel10a,Tel09a,Tel09b,Tel09c,Tel08a}.

In this study, we consider a combined analysis of two different types of signals, hydrogeochemical (sampling frequency of 3 to 6 day$^{-1}$) and geoacoustic (sampling frequency of 1 min$^{-1}$), recorded on the Kamchatka peninsula, Russia.

\section{Nonstationarity factor}

Here, we will only deal with the basic FNS relations needed to understand the nonstationarity factor. The approach is described in detail elsewhere \citep{Tim06a, Tim07a,Tim07b,Tim10a}. The FNS procedures for analyzing original signal $V(t)$, where $t$ is time, are based on the extraction of information contained in autocorrelation function
\begin{equation}
\psi (\tau ) = \left\langle {V(t)V(t + \tau )} \right\rangle, \label{eq1}
\end{equation}
where $\tau$ is the time lag parameter. The angular brackets in relation (\ref{eq1}) stand for the averaging over time interval $T$:  
\begin{equation}
\left\langle {(...)} \right\rangle  = {1 \over T}\int^{T/2}_{-T/2} {(...) \,dt}. \label{eq2}
\end{equation} 

To extract the information contained in $\psi (\tau )$, the following transforms, or "projections", of this function are analyzed: cosine transforms (power spectrum estimates) $S(f)$, where $f$ is the frequency,
\begin{equation}
S(f) = \int^{T/2}_{-T/2} { \left\langle {V(t)V(t + \tau )} \right\rangle \, \cos({2 \pi f t_1}) \,dt_1} \label{eq3}
\end{equation}
and its difference moments (Kolmogorov transient structural functions) of the second order $\Phi^{(2)} (\tau)$
\begin{equation}
\Phi^{(2)} (\tau) = \left\langle {\left[ {V(t) - V(t+\tau )} \right]^2 } \right\rangle. \label{eq4}
\end{equation}

To analyze the effects of nonstationarity in real processes, we study the dynamics of changes in $\Phi^{(2)} (\tau)$ for consecutive "window" intervals [$t_k, t_k+T$], where $k$ = 0, 1, 2, 3, … and $t_k = k \Delta T$, that are shifted within the total time interval $T_{tot}$ of experimental time series ($t_k+T < T_{tot}$). The time intervals $T$ and $\Delta T$ are chosen based on the physical understanding of the problem in view of the suggested characteristic time of the process, which is the most important parameter of system evolution. The phenomenon of "precursor" occurrence is assumed to be related to abrupt changes in functions $\Phi^{(2)} (\tau)$ when the upper bound of the interval  [$t_k, t_k+T$] approaches the time moment $t_c$ of a catastrophic event accompanied by total system reconfiguration on all space scales. 

The analysis of experimental stochastic series often requires the original data to be separated into a smoothed and fluctuation components. In this study, we apply the "relaxation" procedure proposed by \citet{Tim03a} based on the analogy with a finite-difference solution of the diffusion equation, which allows one to split the original signal into low-frequency $V_R(t)$ and high-frequency $V_F(t)$ components. The iterative procedure finding the new values of the signal at every “relaxation” step using its values for the previous step allows one to determine the low-frequency component $V_R(t)$. The high-frequency component $V_F(t)$ is obtained by subtracting $V_R(t)$ from the original signal. This smoothing algorithm progressively reduces the local gradients of the "concentration" variables, causing the points in every triplet to come closer to each other. Such splitting of the original signal $V(t)$ into $V_R(t)$ and $V_F(t)$ makes it possible to evaluate the nonstationarity factor for each of the three functions $V_J(t)$ ($J$ = $R$, $F$, or $G$), where index $G$ corresponds to the original signal.

The FNS nonstationarity factor $C_J (t_k)$ is defined as
\begin{equation}
C_J (t_k) = 2 \times {{{Q_k^J - P_k^J } \over {Q_k^J + P_k^J }}} \times {T \over {\Delta T}}, \label{eq5}
\end{equation}
\begin{equation}
Q_k^J  = {1 \over {\alpha T^2 }}\int\limits_0^{\alpha T} {\,\,\int\limits_{t_k }^{t_k  + T} {\left[ {V_J (t) - V_J (t + \tau )} \right] ^2  dt} \,d\tau }, \label{eq6}
\end{equation}
\begin{equation}
P_k^J  = {1 \over {\alpha T^2 }}\int\limits_0^{\alpha T} {\,\,\int\limits_{t_k }^{t_k  + T - \Delta T} {\left[ {V_J (t) - V_J (t + \tau )} \right] ^2  dt} \,d\tau }. \label{eq7}
\end{equation}						
Here, $J$ indicates which function $V_J(t)$ ($J$ = $R$, $F$ or $G$) is used. Expressions (\ref{eq6}-\ref{eq7}) are given in discrete form elsewhere \citep{Tim10b}. Note that functions $\Phi_J^{(2)} (\tau)$ can be reliably evaluated only on the $\tau$ interval of [0,  $\alpha T$], which is less than half of the averaging interval $T$; i.e., $\alpha < 0.5$.

\section{Experimental data for the Kamchatka peninsula}

The data were recorded in the south-eastern part of the Kamchatka peninsula located at the Russian Far East. The eastern part of the peninsula is one of the most seismically active regions in the world. The area of highest seismicity localized in the depth range between 0 and 40 km represents a narrow stripe with the length of approximately 200 km along the east coast of Kamchatka, which is bounded by a deep-sea trench on the east \citep{Fed85a}.

Specialized measurements of underground water characteristics were started in 1977 to find and study hydrogeochemical precursors of Kamchatka earthquakes. Currently, the observation network includes four stations in the vicinity of Petropavlosk-Kamchatsky (Fig. 1). The Pinachevo station includes five water reservoirs: four warm springs and one borehole GK-1 with the depth of 1,261 m. The Moroznaya station has a single borehole No. 1 with the depth of 600 m. The Khlebozavod station also includes a single borehole G-1 with the depth of 2,540 m, which is located in Petropavlosk-Kamchatsky. The Verkhnyaya Paratunka station comprises four boreholes (GK-5, GK-44, GK-15, and GK-17) with depths in the range from 650 to 1208 m.

The system of hydrogeochemical observations includes the measurement of atmospheric pressure and air temperature, measurement of water discharge and temperature of boreholes and springs, collection of water and gas samples for their further analysis in laboratory environment. For water samples, the following parameters are determined: pH; ion concentrations of chlorine (Cl$^-$), bicarbonate (HCO$_3^{-}$), sulfate (SO$_4^{2-}$), sodium (Na$^+$), potassium (K$^+$), calcium (Ca$^{2+}$), and magnesium (Mg$^{2+}$); concentrations of boric (H$_3$BO$_3$) and silicone (H$_4$SiO$_4$) acids. For the samples of gases dissolved in water, the following concentrations are determined: methane (CH$_4$), nitrogen (N$_2$), oxygen (O$_2$), carbon dioxide (CO$_2$), helium (He), hydrogen (H$_2$), hydrocarbon gases: ethane (C$_2$H$_6$), ethylene (C$_2$H$_4$), propane (C$_3$H$_8$), propylene (C$_3$H$_6$), butane (C$_4$H$_{10}$n), and isobutane (C$_4$H$_{10}$i). The data are recorded at nonuniform sampling intervals with one dominant sampling frequency. For the Pinachevo, Moroznaya, and Khlebozavod stations, this average sampling frequency is one measurement per 3 days; for the Verkhnyaya Paratunka station, one measurement per 6 days. Multiple studies of the hydrogeochemical data and corresponding seismic activity for the Kamchatka peninsula reported anomalous changes in the chemical and/or gas composition of underground waters prior to several large earthquakes in the time frame from 1987 to 2001 \citep{Kop94a,Bel98a,Bia00a,Bia06a,Kha06a}. In this study, we analyze the variations of chlorine-ion concentration determined by a titrimetric method (relative error from 1 to 10$\%$).

Geoacoustic emissions in the frequency range from 25 to 1,400 Hz (0.7 level) have also been recorded in the deep G-1 borehole of the Khlebozavod station under the supervision of V. A. Gavrilov since August, 2000. The data analyzed in this paper were obtained by a geophone with crystal ferromagnetic sensors \citep{Bel00a}. The output signal of such a sensor is proportional to the third derivative of ground displacement, and the gain slope is 60 dB per decade of frequency change. The geophone was set up at the depth of 1,035 m, which is enough to reduce anthropogenic noise levels by more than two orders of magnitude \citep{Gav08a}. The geophone body was fixed inside the borehole casing by a spring. The vertical channel sensitivity of the geophone is 0.15 V $\times $ s$^3$/m. The sensitivity of horizontal channels is 0.60 V $\times $ s$^3$/m. The sensor output signals are separated by third-octave band pass filters into four frequency bands with central frequencies 30, 160, 560, and 1,200 Hz, which is followed by real-time hardware/software signal processing. The value of postprocessed output signal for each channel is proportional to the average value of input signal for one-minute intervals. More detailed description of geoacoustic emission observations and experimental setup for the G-1 borehole is presented elsewhere \citep{Gav08a}.

\section{Results}

To illustrate the nonstationarity factor and proposed phenomenological method, we analyze the hydrogeochemical data for chlorine-ion concentrations at GK-1 (Pinachevo station) and GK-44 (Verkhnyaya Paratunka station) and geoacoustic emissions at the output of geophone vertical frequency channel with the central frequency of 160 Hz (Z160) in G-1 (Khlebozavod station). Chlorine-ion concentration time series for GK-1 (Cl-GK1) was selected because it is characterized by a unidirectional long-period trend without seasonal variations (Fig. 2) and was already treated as a precursory signal due to a gradual chlorine-ion concentration decline down to a local minimum 30 to 60 days before several earthquakes \citep{Kha06a}. On the other hand, chlorine-ion concentration at GK-44 (Cl-GK44) is not considered as a precursory signal because it is dominated by seasonal concentration changes on the background of a slowly varying local mean, the minimum value of which is reached shortly after the strong earthquake on December 5, 1997 ($M_l$ =7.0). The Z160 signal was selected from the whole set of geoacoustic time series because it contains the lowest level of noise (highest signal-to-noise ratio).

To keep the statistical structure of source time series practically intact, the signals were subjected only to minimal preprocessing, which included the removal of single-point spikes, reduction of the hydrogeochemical time series to uniform sampling intervals using linear interpolations, and extraction of every 30$^{\textrm{th}}$ point in the geoacoustic time series to form a new time series with the frequency of 30 min$^{-1}$. Then the time series $V_G(t)$ were separated out into low-frequency $V_R(t)$ and high-frequency $V_F(t)$ components, which were used to calculate the nonstationarity factors. In evaluating $C_J$ ($J$ = $R$, $F$ or $G$) for the hydrogeochemical series, averaging time intervals $T$ in the range from 50 to 900 days were used. For the geoacoustic time series, the interval $T$ was varied from 3 to 20 days. Our analysis showed that the values of $T$ equal to 600 and 20 days are most adequate for locating precursors in the hydrogeochemical and geoacoustic series, respectively. 

Figures 2 and 3 show the variations of $C_J$ for Cl-GK1 and Gl-CK44 together with largest seismic events. It can be seen that spikes in $C_J$ precede several large earthquakes. It should be noted that the low-frequency component $C_R$ shows the most number of precursors for Cl-GK1 and high-frequency component is most informative for Cl-CK44. The first fact is in agreement with the study of \citet{Kha06a}. The second fact implies that the use of the high-frequency component eliminated seasonal changes from the analysis and made Cl-GK44 a precursory signal. Therefore, the FNS nonstationarity factor together with the procedure for separating out high-frequency and low-frequency signal components can be used to analyze different signals despite major differences in their specific features.

Figure 4 shows a combined analysis of hydrogeochemical and geoacoustic variations in the time frame around the October 8, 2001 earthquake ($M_l$ =6.3, $H$ = 24 km, $D$ = 134 km from Petropavlovsk-Kamchatsky), which was the strongest earthquake (based on local magnitude and distance to the epicenter) recorded for the whole time interval of geoacoustic observations in the G-1 borehole. Nonstationarity factors $C_R$ for Cl-GK1 and $C_F$ for Cl-GK44 show spikes with highest values (precursors) in the time frame from 50 to 70 days before the earthquake. $C_G$ for G-1 (the signal is a high-frequency one by its nature) shows precursors 29 and 6 days before the event, which is in agreement with the results reported by \citet{Gav08a}. In other words, anomalous changes in the geoacoustic signal happen closer to the earthquake than in the hydrogeochemical ones, which implies that precursory signals of different nature may take place at different timescales before a large earthquake.

\conclusions
%% \conclusions[modified heading if necessary]
The above example shows that precursory signals of different types may be observed in the same local seismically active zone at different times prior to a large earthquake, which may be attributed to some system preparation preceding the seismic event.  In the studied case, the qualitative changes may be related to a system-wide structural medium reconfiguration at the preparatory phase of the earthquake.

This study also shows that the FNS nonstationarity factor can be used as the standard criterion to detect qualitative changes within relatively short time intervals in virtually arbitrary signals, even if the signals contain strongly pronounced periodic components, as was the case for Cl-CK44. It should be noted that the nonstationarity factor should be analyzed not only for the original signal, but also for its smoothed (low-frequency) and fluctuation (high-frequency) components.

In order to validate the proposed phenomenological systems approach, comprehensive monitoring of seismically active regions such as the Kamchatka peninsula should be performed and the data should be analyzed with the FNS nonstationarity factor. The measured characteristics should include geoacoustic, hydrological/hydrochemical, gasgeochemical, geodetic, and electromagnetic signals summarized in the introduction.

\begin{acknowledgements}
This study was supported in part by the Russian Foundation for Basic Research, project nos. 09-05-98543 and 10-02-01346.
\end{acknowledgements}

%% Literature citations
%% command                        & example result
%% \citet{jones90}|               & Jones et al.\ (1990)
%% \citep{jones90}|               & (Jones et al., 1990)
%% \citep{jones90,jones93}|       & (Jones et al., 1990, 1993)
%% \citep[p.~32]{jones90}|        & (Jones et al., 1990, p.~32)
%% \citep[e.g.,][]{jones90}|      & (e.g., Jones et al., 1990)
%% \citep[e.g.,][p.~32]{jones90}| & (e.g., Jones et al., 1990, p.~32)
%% \citeauthor{jones90}|          & Jones et al.
%% \citeyear{jones90}|            & 1990

\bibliographystyle{copernicus}
\bibliography{nhess}

\newpage

\begin{figure}[t]
\vspace*{2mm}
\begin{center}
\includegraphics[width=8.3cm]{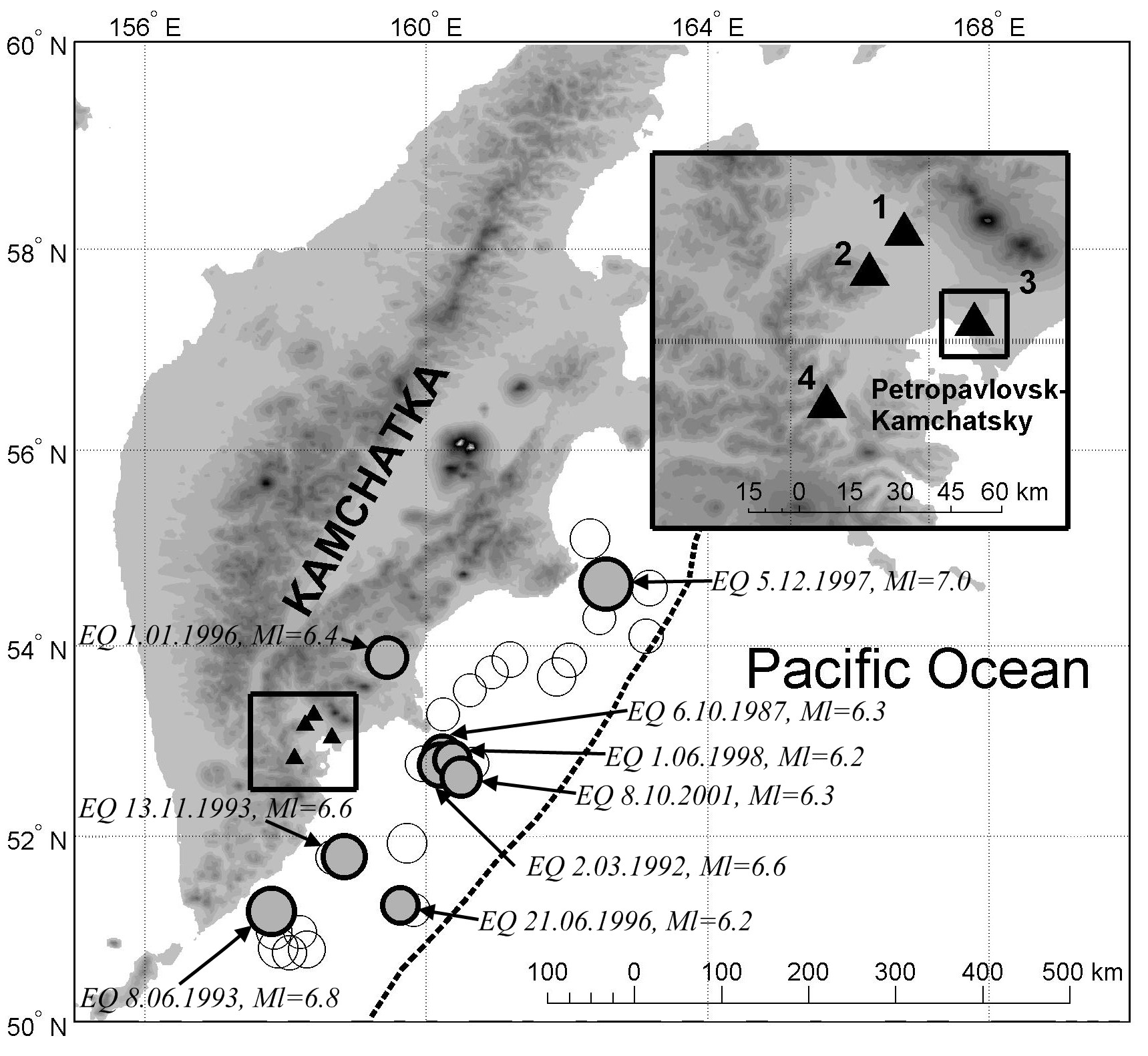}
\end{center}
\caption{Schematic of the measurement area (small rectangular frame on the left) and epicenters of largest earthquakes ($M_l \geq 6$, $H \leq 50$ km, $D \leq$ 350 km) from 1985 to 2009, where $M_l$ -- local earthquake magnitude, $H$ -- depth, $D$ -- distance from the epicenter. The large frame on the right shows a zoomed-in view of the positions of hydrogeological stations: 1 -- Pinachevo, 2 -- Moroznaya, 3 -- Khlebozavod, 4 -- Verknyaya Paratunka. The solid circles denote the earthquakes reportedly preceded by hydrogeochemical anomalies. The dashed line is the axis of the deep-sea trench. The earthquakes were selected using the catalog of Geophysical Survey, Kamchatka Branch of the Russian Academy of Sciences.}
\end{figure}

\begin{figure*}[t]
\vspace*{2mm}
\begin{center}
\includegraphics[width=12cm]{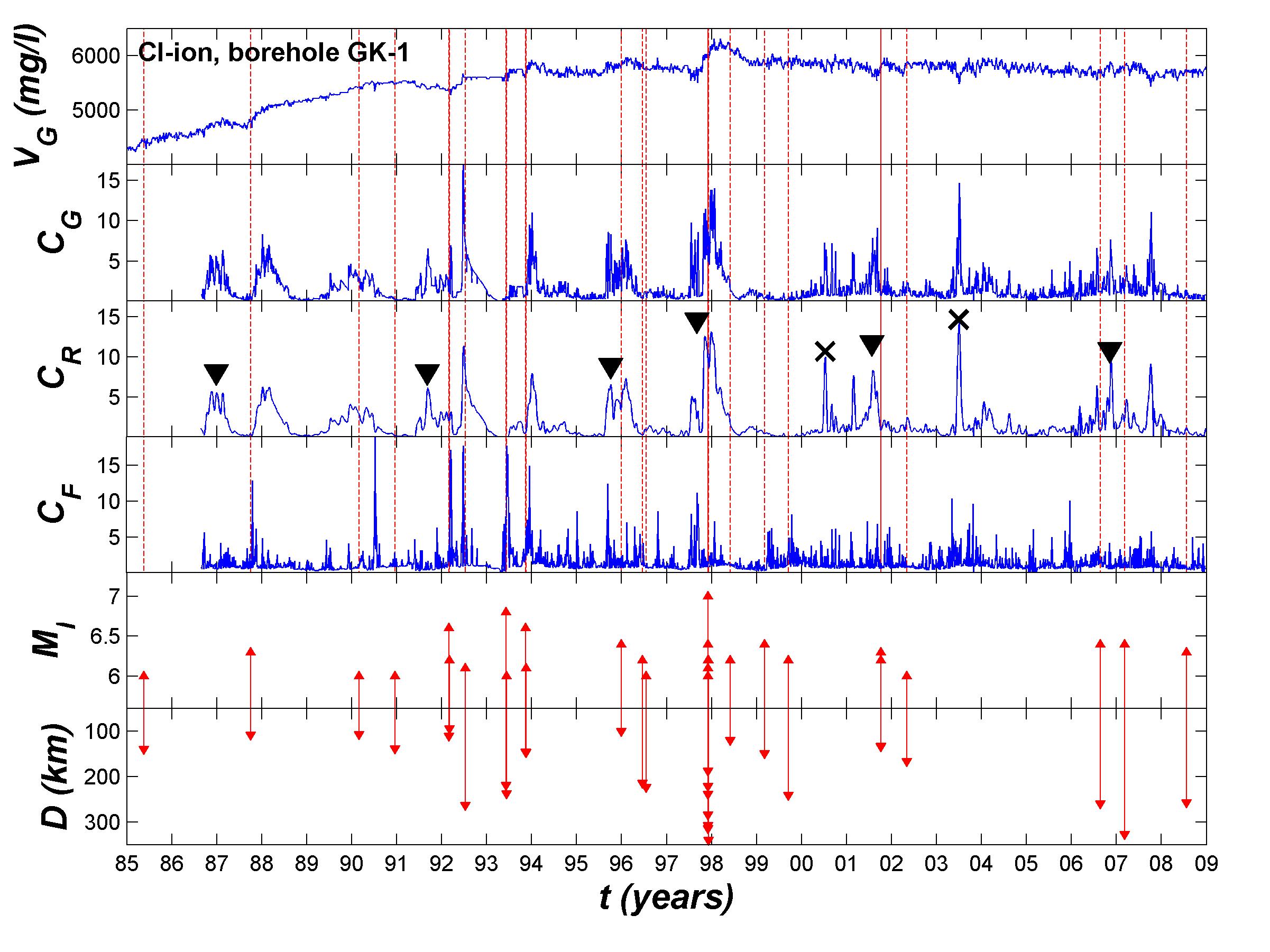}
\end{center}
\caption{Comparison of nonstationarity factor $C_J$ ($T$=600 days, $\Delta T$=3 days) for the GK-1 chlorine-ion concentration time series with seismic activity: $V_G$ -- source signal; $C_G$ -- nonstationarity factor for $V_G$, $C_R$ -- nonstationarity factor for the low-frequency component of $V_G$, $C_F$ -- nonstationarity factor for the high-frequency component of $V_G$, $M_l$ -- local earthquake magnitude, $D$ -- distance from the epicenter. Solid triangles denote sample $C_R$ spikes preceding large earthquakes. Crosses denote sample $C_R$ spikes not related to large seismic events.}
\end{figure*}

\begin{figure*}[t]
\vspace*{2mm}
\begin{center}
\includegraphics[width=12cm]{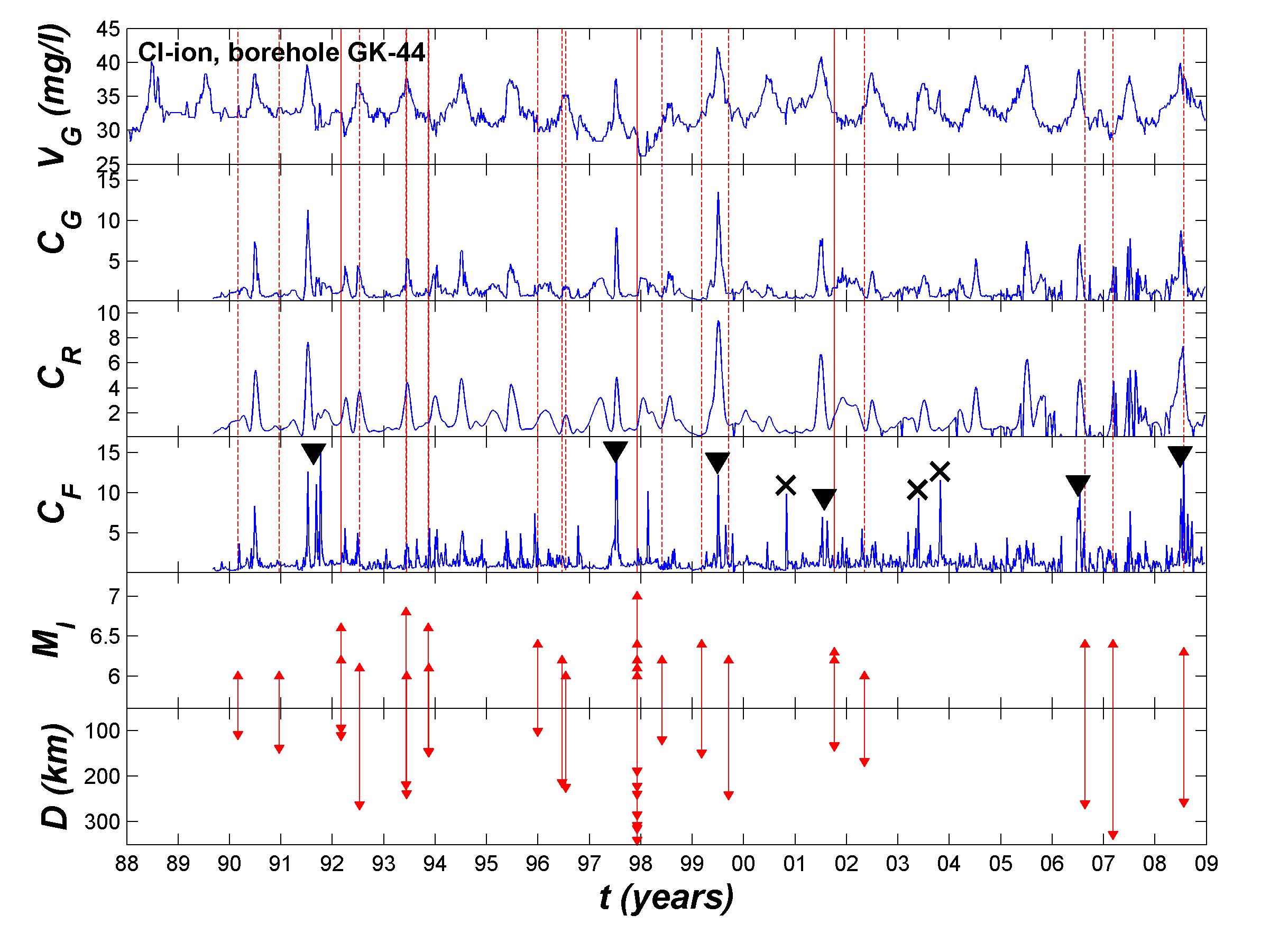}
\end{center}
\caption{Comparison of nonstationarity factor $C_J$ ($T$=600 days, $\Delta T$=3 days) for the GK-44 chlorine-ion concentration time series with seismic activity: Nomenclature as in Fig. 2. Solid triangles denote sample $C_F$ spikes preceding large earthquakes. Crosses denote sample $C_F$ spikes not related to large seismic events.}
\end{figure*}

\begin{figure*}[t]
\vspace*{2mm}
\begin{center}
\includegraphics[width=12cm]{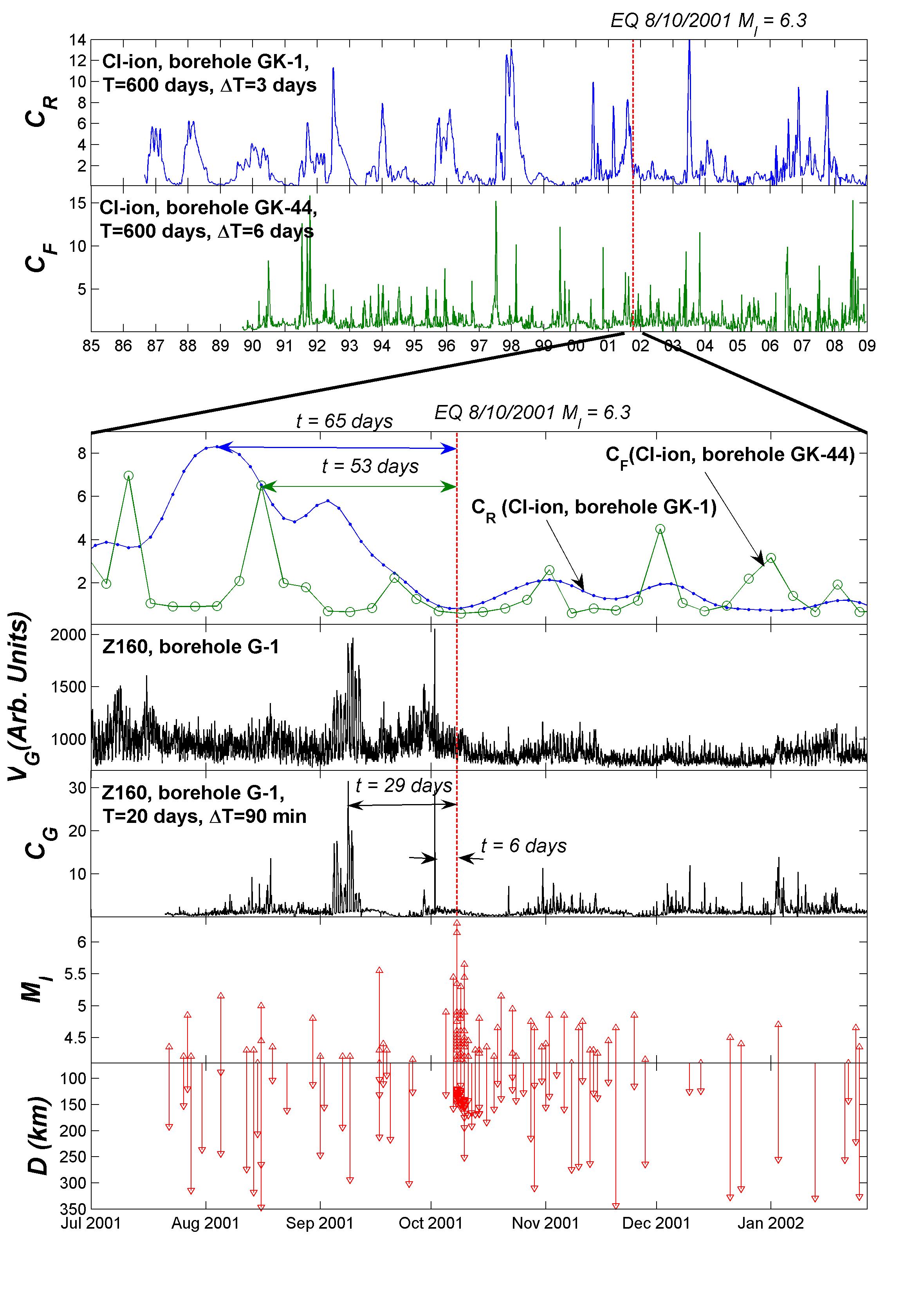}
\end{center}
\caption{Nonstationarity factors for GK-1 and GK-44 chlorine-ion concentrations and Z160 G-1 geoacoustic emissions in the time frame around the 8/10/2001 earthquake. $M_l$ -- local earthquake magnitude, $D$ -- distance from the epicenter. The double-headed arrows denote the time intervals between the nonstationarity factor spikes and earthquake itself.}
\end{figure*}

\end{document}